\documentclass[12pt]{article}

\usepackage{graphicx}
\usepackage{epsfig}
\usepackage{psfrag}
\usepackage{wrapfig}
\usepackage[all]{xy}

\usepackage{fullpage}
\usepackage{setspace}
\usepackage{mathptmx}
\usepackage{url}
\usepackage{booktabs}

\usepackage{footmisc}
\newcommand\wordcount{
    \immediate\write18{texcount -sum -1 \jobname.tex > 'count.txt'}
\input{count.txt}words}

\usepackage{natbib}

\usepackage{amsmath}
\usepackage{amsfonts}
\usepackage{amsthm}
\usepackage{bbm}
\usepackage{array}

\newtheorem{lemma}{Lemma}
\newtheorem{prop}{Proposition}
\newtheorem{corollary}{Corollary}

\def\Var{{\rm Var}\,}
\def\E{{\rm E}\,}

\def\Cov{{\rm Cov}\,}

\newcommand{\X}{\mathbf{X}}

\newcommand{\W}{\mathbf{W}}

\newcommand{\A}{\mathbf{A}}
\newcommand{\B}{\mathbf{B}}

\newcommand{\M}{\mathbf{M}}

\newcommand{\arrowp}{\stackrel{p}{\rightarrow}}

\begin{document}

\sloppy
%\author{anon.}
\author{Peter M. Aronow, Cyrus Samii and Valentina A. Assenova\thanks{Peter M. Aronow is Assistant Professor, Department of Political Science, Yale University, 77 Prospect St., New Haven, CT 06520 (Email: peter.aronow@yale.edu).  Cyrus Samii (contact author) is Assistant Professor, Department of Politics, New York University, 19 West 4th St., New York, NY 10012 (Email: cds2083@nyu.edu). Valentina A. Assenova is Doctoral Student, School of Management (Organizations and Management), Yale University, 165 Whitney Avenue, New Haven, CT 06520 (Email: valentina.assenova@yale.edu). The authors thank Neal Beck, Allison Carnegie, Dean Eckles, Donald Lee, Winston Lin, Kelly Rader, Olav Sorenson, the {\it Political Analysis} editors and two reviewers for helpful comments. We thank Jonathan Baron and Lauren Pinson for research assistance.  Replication materials are available on the {\it Political Analysis} Dataverse (https://dataverse.harvard.edu/dataverse/pan).}}

\title{Cluster-Robust Variance Estimation for Dyadic Data}
\maketitle
\thispagestyle{empty}

\clearpage
\thispagestyle{empty}
\begin{center}
{\LARGE Cluster-Robust Variance Estimation for Dyadic Data}
\end{center}
\vspace{2cm}

\begin{abstract}
Dyadic data are common in the social sciences, although inference for such settings involves accounting for a complex clustering structure. Many analyses in the social sciences fail to account for the fact that multiple dyads share a member, and that errors are thus likely correlated across these dyads. We propose a nonparametric, sandwich-type robust variance estimator for linear regression to account for such clustering in dyadic data. We enumerate conditions for estimator consistency. We also extend our results to repeated and weighted observations, including directed dyads and longitudinal data, and provide an implementation for generalized linear models such as logistic regression. We examine empirical performance with simulations and an application to interstate disputes.
\end{abstract}

\vspace{2cm}

\noindent Keywords: cluster robust variance estimation, dyadic data, agnostic regression\\
\vspace{.05in}\\
\noindent Word count: 5,022 (including main text, appendix for print, captions, and references).

\doublespace

\clearpage
\setcounter{page}{1}

\section{Introduction}

Dyadic data are central in social science applications ranging from international relations to ``speed dating.''\footnote{For example, in the past five years the {\it American Political Science Review}, {\it American Journal of Political Science}, and {\it International Organization} have together published 62 papers that include dyadic analyses.} A challenge with dyadic data is to account for the complex dependency structure that arises due to the connections between dyad members. For instance, in a study of international conflict, a change in leadership in one country may affect relations with all countries with which that country is paired in the data, thereby inducing a correlation between these dyadic observations.  This generates a cluster of dependent observations associated with that country.  As leadership changes occur in multiple countries, the correlations emanating from each of these countries will overlap into a web of interwoven clusters.  We refer to such interwoven dependency in dyadic data as  ``dyadic clustering.'' By ignoring the dyadic clustering, the analysis would take the dyad-level changes emanating from a single leadership change as independently informative events, rather than a single, clustered event.  An analysis that only accounts for correlations in repeated observations of dyads (whether by clustering standard errors or using dyad fixed effects) would fail to account for such inter-dyad correlation.  

Statistical analysis of dyads typically estimates how dyad-level outcomes (e.g., whether there is open conflict between countries or the decision for one person to ask another on a date) relate to characteristics of the individual units as well as to the dyad as a whole (e.g., measures of proximity between units). The usual approach is to regress the dyad-level outcome on unit- and dyad-level predictors.  Due to dyadic clustering, the observations contributing to such an analysis are not independent.  Failure to account for dyadic clustering may result in significance tests or confidence intervals with sizes that are far too small relative to the true distribution of the parameters of interest.

We establish sufficient conditions for the consistency of a non-parametric sandwich estimator \citep{huber1967-robust, white80} for the variance of regression coefficients under dyadic clustering. Cluster-robust sandwich estimators are common for addressing dependent data (\citealp[Ch. 8]{angrist_pischke09}; \citealp{liang_zeger86_gee}). \citet{cameron_etal2011_multiway} provide a sandwich estimator for ``multi-way'' clustering, accounting, for example, for clustering between people by geographic location and age category. We extend these methods to dyadic clustering, accounting for the fact that dyadic clustering does not decompose neatly into a few crosscutting and disjoint groups of units; rather, each {\it unit} is the basis of its own cluster that intersects with other units' clusters. \citet[Eq. 2.5]{fafchamps-gubert2007-risk-networks} propose a sandwich estimator for dyadic clustering that is very similar to what we propose below. Their derivation is constructed through analogy to the results of \citet{conley99_spatial}. However, neither paper establishes conditions for consistency under dyadic clustering. We establish such consistency conditions. We also provide extensions to the longitudinal case and generalized linear models such as logistic regression. We evaluate performance with simulations and a reanalysis of a classic study on interstate disputes \citep{russett-oneal2001-triangle-book}. The appendix generalizes to weighted data and generalized linear models, and the supporting information provides another illustration with a speed dating experiment \citep{Fisman2006}. 

Current statistical approaches to handling dyadic clustering include the use of parametric restrictions in mixed-effects models \citep{gelman-hill2007-multilevel-book, hoff2005-dyad-mixed-effects, kenny-etal2006-dyadic}, the spatial error-lag model \citep{beck-etal2006-spatial-lags}, or permutation inference for testing against sharp null hypotheses \citep{erikson-etal2014-permutation-dyads}. These approaches have important limitations that our approach overcomes. First, mixed effects models and spatial lag models impose a parametric structure to address the clustering problem. This makes them sensitive to misspecification of the conditional mean (that is,  deviations between the data and the linearity assumption). Our approach is robust in that it provides asymptotically valid standard errors even under such misspecification.  This is valuable in itself to the extent that models are typically approximations \citep{buja-etal2014-models-as-approximations, hubbard-etal2010-to-gee}, although readers should not take this to mean that they are free from the obligation to fit as good an approximation as possible.  It is also valuable in providing a reliable benchmark to use in evaluating model specification \citep{king-roberts2014-robust, white80, white-1981-detection-misspec}.  Second, \citet{erikson-etal2014-permutation-dyads}'s solution of non-parametric randomization inference does not provide a procedure for obtaining valid confidence intervals, while our procedure does. Third, all three of these alternatives require considerable computational power, possibly exceeding available computational resources even for data analyses that are common in international relations or network analysis. We demonstrate below that a mixed-effects or spatial error-lag approach is infeasible in a typical international relations example.  In contrast, our proposed estimator is easy to compute. The variance estimation methods we develop are a natural complement to non- and semi-parametric approaches to regression with dyadic data \citep{green_etal01_dirty_pool}.

\section{Setting}

We work within the ``agnostic" \citep{angrist-imbens2002-agnostic, lin11_freedmans_critique} or ``assumption lean'' \citep{buja-etal2014-models-as-approximations} regression framework developed by \citet{angrist_pischke09}, \citet{goldberger1991_course}, and \citet{white-1980-unknown, white-1981-mle-misspec}.\footnote{Our results would  hold under a regression model approach along the lines of \cite{greene08_ec_analysis} or \citet{davidson_mackinnon04_book}. The value of using the ``agnostic regression'' approach is that the derivation is robust to violations of the assumption that the regression specification is correct for the conditional expectation. Nevertheless, this should not be taken as an endorsement for ignoring misspecification or assuming that linear approximations with robust standard errors can cure misspecification problems \citep{king-roberts2014-robust}. This paper leaves aside issues of causal identification and focuses only on estimation and statistical inference.  Nonetheless, the issues discussed here would be relevant for studies that attempt to estimate causal effects with dyadic data.} We begin with a cross-section of undirected dyads and derive the basic convergence results for this case. Below we extend these results to repeated dyads, which covers directed dyads and longitudinal data.  Proofs are in the appendix.

Begin with a large population from which we take a
%i.i.d. 
random (i.i.d.)
sample of units, with the sample members indexed by $i=1,...,N$ and grouped into $D = {N \choose 2} = N(N-1)/2$ dyads. Pairs of unit indices within each dyad map to dyadic indices, $1,...,D$, through the index function $d(i,j) = d(j,i)$, with an inverse correspondence $s(d(i,j)) = \{i,j\}$, and we assume no $d(i,i)$ type dyads. Consider a linear regression of $Y_{d(i,j)}$ on a $k$-length column vector of regressors $X_{d(i,j)}$ (which includes the constant):
$$
Y_{d(i,j)} = X_{d(i,j)}'\beta + \epsilon_{d(i,j)},
$$
where $\beta$ is the slope that we would obtain if we could fit this model to the entire population, allowing for possible non-linearity in the true relationship between $\E[Y_{d(i,j)}|X_{d(i,j)}]$ and $X_{d(i,j)}$, and  $\epsilon_{d(i,j)}$ is the corresponding population residual.\footnote{If $\E[Y_{d(i,j)}|X_{d(i,j)}]$ is nonlinear in $X_{d(i,j)}$, the population residual will itself vary in $X_{d(i,j)}$, which  undermines inference that assumes that $\epsilon_{d(i,j)}$ is independent of $X_{d(i,j)}$.  The agnostic regression approach avoids such an assumption.} To lighten notation, for the remainder of the discussion we use $d = d(i,j)$.

Define the sample data as $\X = (X_1...X_D)'$ and $Y = (Y_1 ...Y_D)'$. The ordinary least squares (OLS) estimator is
$$
\hat \beta = (\X'\X)^{-1}\X'Y,
$$ 
with residual $e_d = Y_d - X_d'\hat \beta$. Since the values of $Y_{d}$ and $X_{d}$ are determined by the characteristics of the units $i$ and $j$, $\epsilon_d$ and $\epsilon_{d'}$ for dyads containing either unit $i$ or $j$ are allowed to be correlated by construction. However, by random
% via i.i.d.
sampling of units, $\Cov(\epsilon_{d},\epsilon_{d'}) = 0$ for all dyads that do not share a member. The number of pairwise dependencies between $i$ and $j$ for which $\Cov(\epsilon_{d},\epsilon_{d'}) \neq 0$ is $O(N^3)$. Let the support for $X_d$ and $Y_d$ be bounded in a %positive, 
finite interval and assume the usual rank conditions on $\X$ \citep{chamberlain1982-regression-panel, white-1984-asymptotics}.  

\begin{lemma}\label{lemma-var}
Suppose data and $\hat \beta$ as defined above.  As $N \rightarrow \infty$, the asymptotic distribution of  $\sqrt{N}(\hat \beta - \beta)$ has mean zero and variance
\begin{equation}
V = \frac{N}{D^2}\E[X_dX_d']^{-1}\Var\left[\sum_{d=1}^D X_d\epsilon_d\right]\E[X_dX_d']^{-1}, \label{varB}
\end{equation}
with,
\begin{equation}
\Var\left[\sum_{d=1}^D X_d\epsilon_d \right] = \sum_{d=1}^D \left( \underbrace{\E\{X_{d}X'_d \Var[\epsilon_d|\X]\}}_{A} + \underbrace{\sum_{d' \in \mathcal{S}(d)} \E\{X_{d}X'_{d'} \Cov[\epsilon_d,\epsilon_{d'}|\X]\}}_{B} \right), \label{varxe}
\end{equation}
where $\mathcal{S}(d) = \{d' \ne d \} \cap \{d': s(d)\cap s(d') \ne \emptyset \}$, the set of dyads other than $d$ that share a member from $d$.
\end{lemma}
$A$ is the dyad-specific contribution to the variance, and $B$ is the contribution due to inclusion of common units in multiple dyads. Note that these features of the distribution of $\sqrt{N}(\hat \beta - \beta)$ establish the consistency of $\hat \beta$ as well, which is not surprising given standard results for the consistency of OLS coefficients on dependent data \citep{white-1984-asymptotics}.

\section{Identification and Estimation}

Given data as defined above, we examine the properties of a plug-in variance estimator that is analogous to the sandwich estimators defined for heteroskedastic or clustered data 
\citep{arellano87-cluster-robust, cameron_etal2011_multiway, liang_zeger86_gee, white80}. We establish sufficient conditions for consistency of the plug-in variance estimator. We consider the cross-sectional case and repeated observations case. The appendix also contains extensions to weighted observations and generalized linear models.

\begin{prop}
Define the variance estimator
\begin{equation}
\hat V = (\X'\X)^{-1}\sum_{d=1}^D \left(X_{d}X'_d e_d^2 + \sum_{d' \in \mathcal{S}(d)} X_{d}X'_{d'} e_d e_{d'} \right)(\X'\X)^{-1}.
\end{equation}
Under the conditions of Lemma \ref{lemma-var} and if $X_d$ and $e_d$ have bounded support with non-zero second moments and cross-moments, then as $N \rightarrow \infty$, $${N \hat V} - V \arrowp 0.$$
\label{prop1}
\end{prop}
The proposition indicates that $\hat V$ provides a consistent estimator to characterize the true asymptotic sampling distribution of the regression coefficients, $\hat \beta$.  Standard error estimates for $\hat \beta$ are obtained from the square roots of the diagonals of $\hat V$.  The assumption of bounded support for $X_d$ and $e_d$ merely rules out situations that are unlikely to arise in real world data anyway (e.g., where the mass of the data pushes out toward infinity as $N$ grows).

Repeated dyad observations are common in applied settings. For example, the data may include multiple observations for dyads over time.  Dyadic panels are very common in studies of international relations.  
Or, if the dyadic information is directional, then the data will contain
two observations for each dyad, with one observation capturing outcomes that go in the $i$
to $j$ direction, and the other capturing outcomes that go in the $j$ to $i$
direction.  This is conceptually distinct from repeated observations over time. But if there are dyadic dependences for both senders and receivers of a directed dyad, then the dependence structure for a pair of directed $i$-$j$ dyads will be the same as if we had repeated observations of an undirected $i$-$j$ dyad.  The results above translate straightforwardly to the repeated dyads
setting. Formally, suppose that for each dyad $d$ observations are
indexed by $t=1,...,T(d)$, where the $T(d)$ values are fixed and finite. Let $(Y_{dt},X_{dt}')$ represent the data for observation $t$ from dyad $d$, 
$$
\begin{array}{ccc} Y_d = \left( \begin{array}{c} Y_{d1} \\ \vdots \\ Y_{dT(d)} \end{array}\right) & \text{and} & X_d = \left( \begin{array}{c} X_{d1}' \\ \vdots \\ X_{dT(d)}' \end{array}\right), \end{array}
$$
and $Y = (Y_1' \hdots Y_d' \hdots Y_D')'$ and $\X = (X_1' \hdots X_d' \hdots X_D')'$, that is, the stacked $Y_d$ vectors and $X_d$ matrices.  Let $\beta_r$ and $\hat
\beta_r$ denote, respectively, the population slope and OLS estimator as defined above but now applied to the repeated
dyads data.

\begin{corollary}
For the repeated dyads case, assume the same conditions as in Proposition \ref{prop1} and consider the following variance estimator
$$
\hat V_r = (\X'\X)^{-1}\sum_{d=1}^D\left(X_d'e_de_d'X_d + \sum_{d' \in \mathcal{S}(d)} X_d'e_de_{d'}'X_{d'} \right) (\X'\X)^{-1}.
$$
Then as $N \rightarrow \infty$, 
$$
N\Var[\hat \beta_r - \beta_r] - V_r \overset{p}{\rightarrow} 0,$$
and
$$
N\hat V_r - V_r \overset{p}{\rightarrow} 0,
$$
where 
$$
V_r = \frac{N}{\left[\sum_{d=1}^D T(d)\right]^2} \E[X_dX_d']^{-1}
\E\left( \E\left[X_d'\Cov[\epsilon_d|\X]X_d\right] + \sum_{d' \in \mathcal{S}(d)}\E\left[ X_d'\Cov[\epsilon_d, \epsilon_{d'}|\X]X_{d'}\right]\right)
\E[X_dX_d']^{-1}.
$$
\end{corollary}

Some remarks are in order when it comes to the repeated observations case.  First, our analysis of the repeated observations case takes the $T(d)$ values to be fixed and finite and demonstrates consistency as $N$ grows.  In a cross-national study, this would mean that one should pay most attention to the number of countries, rather than the amount of time, that are in the data.  Under fixed $N$ and growing $T(d)$ further assumptions would be required for consistency, such as serial correlation of fixed order  or, more generally, strong mixing over time.  Second, with repeated observations, a common strategy for identifying effects  is to use dyad-specific fixed effects \citep{green_etal01_dirty_pool}.    With fixed and finite $T(d)$, this does not introduce any new complications.  As in \citet{arellano87-cluster-robust} for the case of serial correlation with fixed effects, the results for dyadic clustering translate directly to centered data, and dyadic fixed-effects regression amounts to centering the data on dyad-specific means.  With fixed $N$ and growing $T(d)$, however, additional assumptions are needed for consistency \citep{hansen07_cluster_robust, stock-watson2008-FE-robust}. 

For efficient computation, we follow \citet{cameron_etal2011_multiway} to perform a ``multi-way decomposition'' of the dyadic clustering structure:
\begin{prop}
We have the following algebraic equivalence,
$$
\hat V_r = \sum_{i=1}^N \hat V_{C,i} - \hat V_D - (N-2) \hat V_{0},
$$
where 
$$
\hat V_{C,i} = (\X'\X)^{-1}\hat \Sigma_{C,i}(\X'\X)^{-1}
$$
$$
\hat V_D = (\X'\X)^{-1} \hat \Sigma_D (\X'\X)^{-1}
$$
$$
\hat V_0 = (\X'\X)^{-1} \hat \Sigma_0(\X'\X)^{-1},
$$
and
$$
\hat \Sigma_{C,i} = \sum_{j\ne i} \sum_{k \ne i} X'_{d(i,j)}e_{d(i,j)}e'_{d(i,k)}X_{d(i,k)} + 
\sum_{j,k \ne i} \sum_{t=1}^{T(d(j,k))} X_{d(j,k)t}X'_{d(j,k)t}e^2_{d(j,k)t},
$$
$$
\hat \Sigma_D = \sum_{d=1}^D X'_de_de'_dX_d, \text{ \hspace{1em} and \hspace{1em}} \hat \Sigma_0 = \sum_{d=1}^D \sum_{t=1}^{T(d)} X_{dt}X_{dt}'e^2_{dt}.
$$
\label{prop-decomp}
\end{prop}
$\hat V_{C,i}$ is the usual asymptotically consistent cluster-robust variance estimator (with no degress-of-freedom adjustment) that clusters all dyads containing unit $i$ and assumes all other observations to be independent. $\hat V_D$ is the same cluster robust estimator but clustering all repeated dyad observations. $\hat V_0$ is the usual asymptotically consistent heteroskedasticity robust (HC) variance estimator that assumes all observations (even within repeated dyad groupings) are independent.  To understand this decomposition, note that dyadic clustering involves clustering on each of $N$ units.  But in summing the contributions from unit-specific clusters (the $\hat V_{C,i}$s), we double count the dyad contributions ($\hat V_D$) and add in the independent contributions ($V_0$) $N$ times.  We can correct for this by subtracting $\hat V_D$, which also removes the $V_0$ component, and then subtracting $N-2$ of the $V_0$ components.  (The cross section cases simply sets $\hat V_D = \hat V_0$). This decomposition shows that one can compute the dyadic cluster robust estimator using readily available robust standard error commands.\footnote{Code for implementation in R is available from the replication archive posted to the Dataverse page references in the acknowledgments note).}  

Our dyadic cluster robust variance estimator allows one to perform valid inference under less restrictive dependence assumptions that are used to identify random effects or spatial error lag models.  Our multi-way decomposition also shows that it is computationally much simpler than those alternatives.  The latter point is relevant when one has many units or time periods in which case random effects and spatial error lag models, despite their restrictions, may be infeasible with current computational resources given the challenges of evaluating a likelihood with as many dependencies as emerge in a dyadic analysis.

\section{Simulation Evidence}

We use Monte Carlo simulations to evaluate the finite sample properties of the proposed estimators under the cross-sectional and repeated dyads settings. We suppose that population values obey the following,
\begin{equation}
Y_{d(i,j)t} = \beta_0 + \beta_1 |X_i - X_j| + \underbrace{\alpha_i + \alpha_j + \nu_{d(i,j)t}}_{\epsilon_{d(i,j)t}}, \label{eq:dgp}
\end{equation}
where $Y_{d(i,j)t}$ is the $t^{th}$ observed outcome for the dyad that includes units $i$ and $j$, $X_i$, $X_j$, $\alpha_i$, $\alpha_j$, and $u_{d(i,j)t}$ are independent draws from standard Normal distributions, and the compound error, $\epsilon_{d(i,j)t}= \alpha_i + \alpha_j + \nu_{d(i,j)t}$, is unobserved. In the cross-sectional case, we only observe one outcome per dyad, so $t=1$ for all observations (that is, the $t$ subscript is extraneous for the cross-sectional case). In the repeated observations case, we have two observations per dyad, so $t=1,2$ for all dyads. We fix $\beta_0=0$ and $\beta_1=1$. We use OLS to estimate $\hat \beta_0$ and $\hat \beta_1$. In the supplemental information we show results for mixed effects models as well. The fact that $X_i$ and $\alpha_i$ are constant across dyads that include unit $i$ (same for $j$) implies non-zero intra-class correlation in both $X$ and $\epsilon$ among sets of dependent dyads, in which case ignoring the dependence structure will tend to understate the variability in $\hat \beta_0$ and $\hat \beta_1$. This is the dyadic version of \cite{moulton86}'s problem. 

\begin{figure}[!hp]
\centering
\includegraphics[width=1\textwidth]{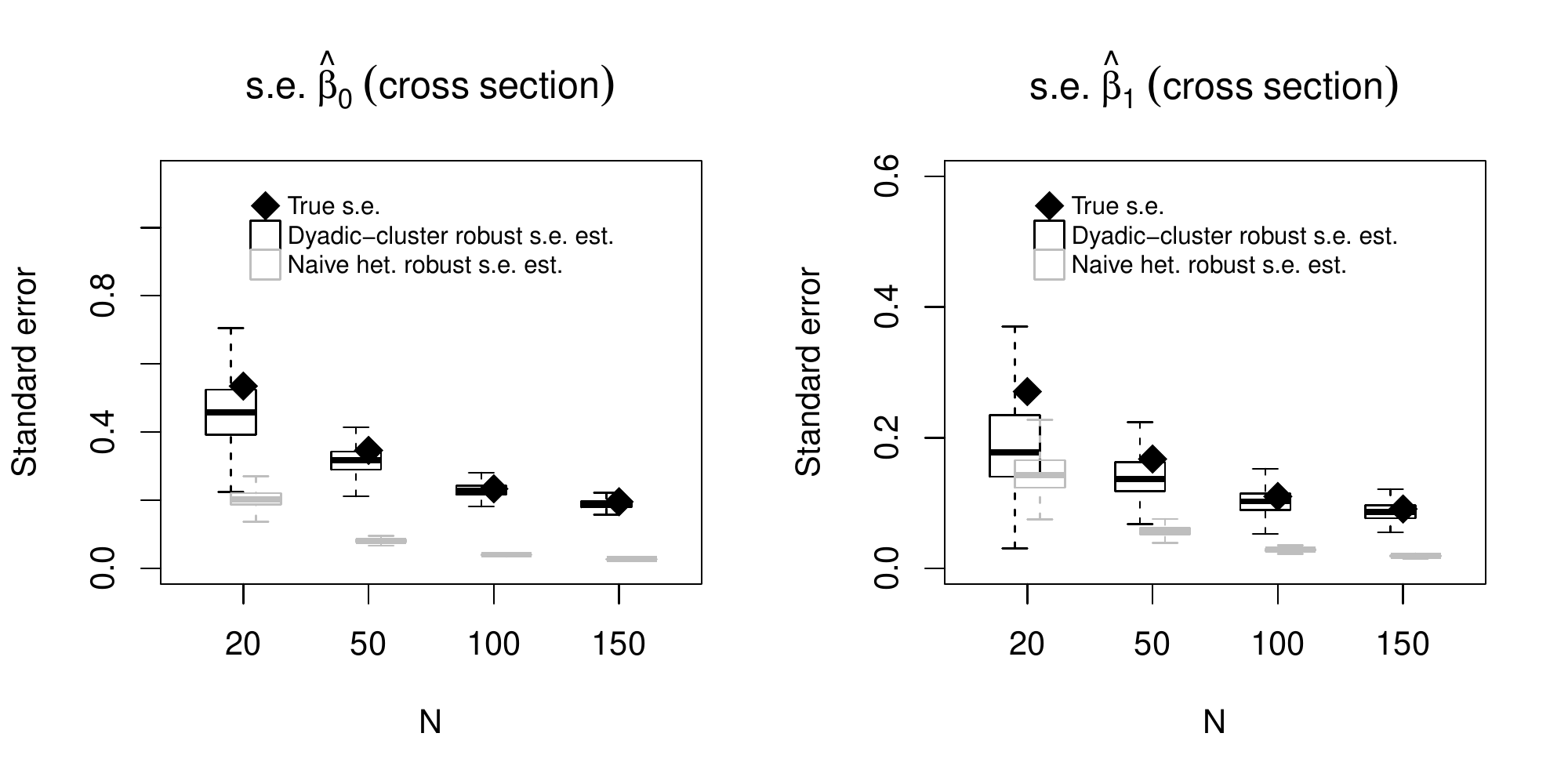}
\includegraphics[width=1\textwidth]{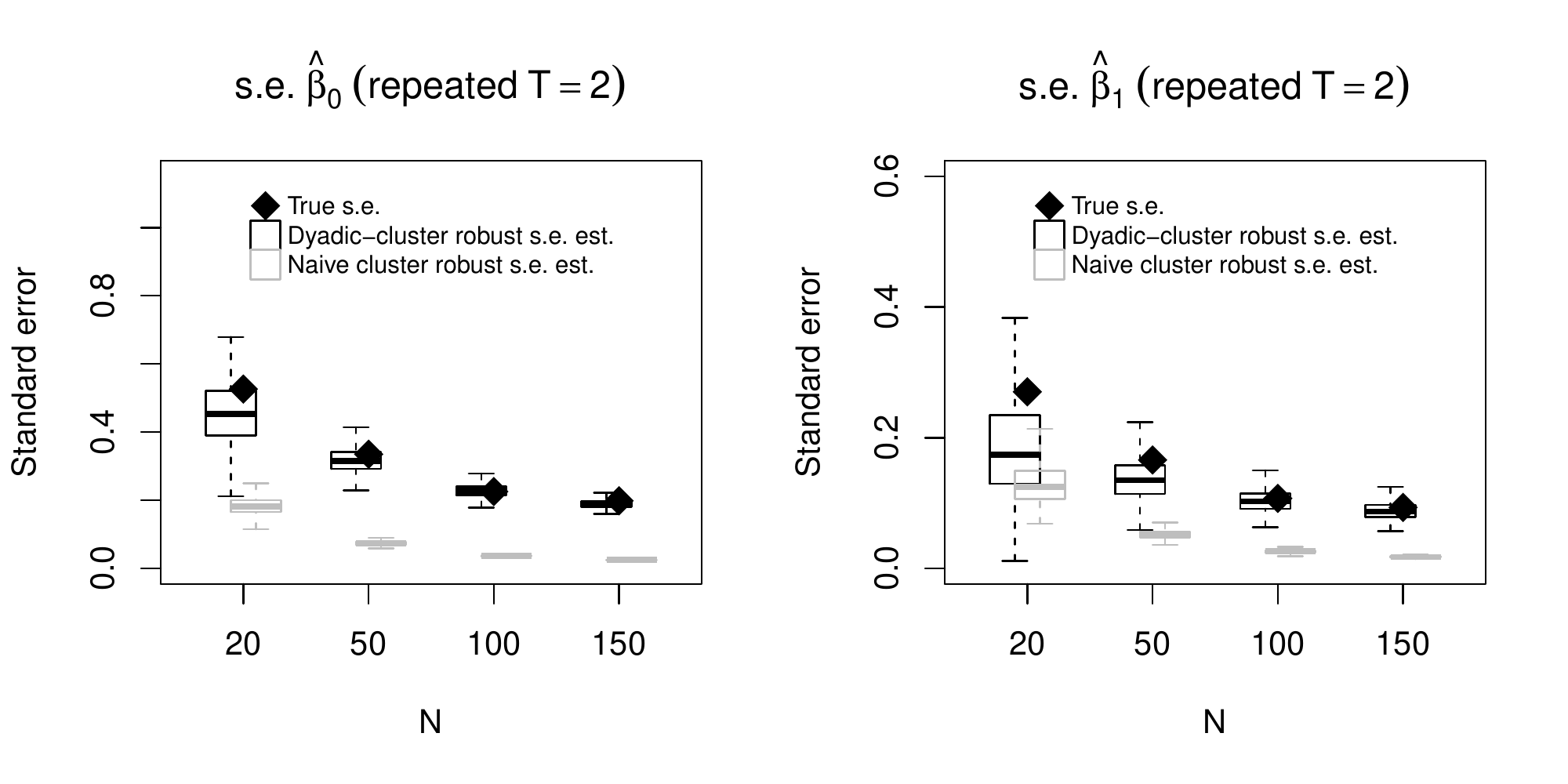}
\caption{Convergence of the standard error estimator as $N$ goes from 20 to 150 (implying number of dyads goes from 190 to 11,175) for a cross section of dyads (top) and repeated dyad observations (T=2 for all dyads). Black box plots show the distribution of standard error estimates using the proposed dyadic cluster robust estimator. In the top two figures, gray box plots show standard error estimates from a heteroskedasticity robust estimator while in the bottom two figures, gray box plots show standard error estimates from a ``naive'' cluster-robust estimator that clusters by dyad over repeated observations (but does not cluster by units across dyads). }
\label{fig:sim1}
\end{figure}

Results from 500 simulation runs are shown in Figure \ref{fig:sim1}.  The $x$-axis shows the sample size, where we show results for samples with 20, 50, 100, and 150 units (implying 190, 1,225,  4,950, and 11,175 dyads, respectively). The $y$-axis is on the scale of the standard error of the coefficients.  The black diamonds plot the empirical standard standard errors for each of the sample sizes (that is, the standard deviation of the simulation replicates of the coefficient estimates, $\hat \beta_0$ and $\hat \beta_1$).  The black box plots show the distribution of our proposed standard error estimator.  The gray box plots show, in the top figures, the distribution of the ``HC2'' heteroskedasticity robust variance estimator, which does not account for either dyadic or repeat observation clustering \citep{white80, mackinnon_white1985_hc2}, and for the bottom figures, the ``cluster robust'' analog of \citet{white80}'s estimator accounting for dependence in repeated dyad observations \citep{arellano87-cluster-robust, liang_zeger86_gee}.  

Convergence of our proposed standard errors based on $\hat V$ and $\hat V_r$ (black box plots) to the true standard error (black diamonds) is quick. The alternative estimators, which do not account for dyadic clustering, grossly understate the variability in the coefficient estimates.  These results suggest that in finite samples from data generating processes that resemble our simulations, the proposed estimator is quite reliable for inference on OLS coefficients so long as the sample size is on the order of 50 to 100 units.  Most applications in political science operate with samples of at least this size.  Changes to the shape of the error distributions (e.g., uniform or bimodal) yield the same results.\footnote{The simulation code, which allows one to choose arbitrary data distributions, is available from the replication archive posted to the Dataverse page referenced in the acknowledgement note.} 

In the supporting information, we demonstrate comparisons between mixed effects models and our ``robust'' estimator.\footnote{\label{fn:sel-model-probs} We sought also to include a comparison to spatial error lag models, but the models failed to converge with $N \ge 100$.  We note that in the original \citet{beck-etal2006-spatial-lags} paper, they limited their analysis of dyadic data to small cross sections.  Our experience suggests that it may be infeasible to fit such models with substantially larger datasets.}  Specifically, we compare OLS with dyadic cluster robust standard errors to a two-way random effects model that incorporates normal random effects for each of the $N$ units \citep{gelman-hill2007-multilevel-book}.  The data generating process captured above (that is, expression \ref{eq:dgp} with normal data) corresponds precisely to the assumptions of a two-way normal random effects model, and so it is no surprise that we find this model to be more efficient than OLS and to produce consistent standard errors.  However, as discussed above, such models are sensitive to misspecification.  For example, in fitting a linear approximation model to mildly non-linear data, the dyadic cluster robust standard errors continue to be consistent for the coefficients of the linear approximation.  The degree of misspecification in this case is not extreme (as illustrated in the appendix) and resembles the kind of approximations that is common in applied work \citep{buja-etal2014-models-as-approximations}. Nonetheless, the random effects standard errors are inconsistent and substantially anti-conservative.  Such sensitivity to misspecification is one problem for the random effects model.  Another problem arises when there is unobserved country-level heterogeneity that could confound estimates of the coefficients of interest.  In such situations, our dyadic cluster robust estimator would be a natural complement to a fixed effects analysis, analogous to what \citet{arellano87-cluster-robust} proposes for non-dyadic repeated observation settings. The third problem arises in computational feasibility, an issue that arises in our application to which we now turn.

\section{International Militarized Disputes Application}

Our application is based on the classic study by \citet{russett-oneal2001-triangle-book} on the determinants of international militarized conflicts.  They study how the likelihood of a militarized dispute between states in a dyad relates to various dyad-level attributes, including an indicator of whether the two states are formal allies, the log of the ratio of an index of military capabilities, the lowest score in the dyad on a democracy index, the ratio of the value of trade and the larger GDP of the two states, an indicator of whether both states are members of a common international organization, an indicator of whether the states are geographically noncontiguous, the log of the geographic (Euclidean) distance between the two states' capitals, and an indicator of whether both states are ``minor powers'' in the international system.  Dyadic clustering could arise in many ways with these data, for example if a country entered into an alliance, thereby changing the joint alliance indicators, or if the military capabilities of a country changed, thereby changing the power ratios.  

We replicate Russett and Oneal's primary analysis as reported in their Table A5.1. They use annual data on 146 states in the international system paired into 1,178 dyads (out of 10,585 possible) and observed for as few as one and as many as ninety years between 1885 and 1991, for a total of 39,988 observations.  In their original analysis, Russett and Oneal fit a GEE model assuming AR(1) errors within dyads over time \citep{zorn2001-gee}.  Table \ref{tab:russett} shows the results of our reanalysis. Columns (1) and (2) replicate the original published results.  Columns (3)-(6) show coefficients and various standard error estimates for a simple (pooled) logistic regression.  There is little difference in the coefficient estimates from the GEE-AR(1) model as compared to the simple logistic regression, so we focus on the standard error estimates.  Column (4) contains estimates that account only for dyad-year heteroskedasticity, column (5) also accounts for arbitrary dependence over time for each dyad, and then column (6) also accounts for dyadic clustering.  Accounting for the dyadic and repeated-observation clustering results in standard error estimates that are sometimes an order of magnitude larger than what we obtain when we ignore all clustering and also considerably larger than what one would estimate were one to account for repeated dyads clustering but ignore the inter-dyadic clustering.  The latter result is also relevant when comparing the standard errors from the original GEE-AR(1) model, as they resemble the estimates in column (5).  In their original analysis, Russett and Oneal found that all of the predictors had a statistically significant relationship to the likelihood of militarized conflict.  But when one takes into account dyadic clustering, the coefficient for international organizations would fail to pass a conventional significance test (e.g., $t > 1.64$). 

\begin{table}[!t]
\caption{Estimates and robust standard error estimates from logistic regression of an indicator for militarized conflict between dyads on dyad-level predictors, 1885-1991, based on \citet[Table A5.1]{russett-oneal2001-triangle-book}.} 
\centering
\begin{footnotesize}
\begin{tabular}{|r|rr|rrrr|}
\hline
& (1) & (2) & (3) & (4) & (5) & (6)\\
& \multicolumn{2}{c|}{Original GEE estimates}& \multicolumn{4}{c|}{Simple logistic regression estimates} \\
& &  & & Het. & Naive Cluster& Dyadic Cluster \\
Predictor of Militarized Conflict & Coef. & S.E. & Coef. & Robust S.E. & Robust S.E. &  Robust S.E. \\
\hline
Alliances  					& -0.539 & 0.159 & -0.595  & 0.069  & 0.175  & 0.265 \\
Power ratio 				& -0.318 & 0.043 & -0.328  & 0.020  & 0.047  & 0.070 \\
Lower democracy  			& -0.061 & 0.009 &  -0.064 & 0.005  & 0.010  & 0.015 \\
Lower dependence 			& -52.924& 13.405& -67.668 & 10.734 & 17.560 & 24.749 \\
International organizations & -0.013 & 0.004 & -0.011  & 0.002  & 0.005  & 0.008 \\
Noncontiguity 				& -0.989 & 0.168 &  -1.048 & 0.074  & 0.181  & 0.185 \\
Log distance 				& -0.376 & 0.065 & -0.375  & 0.026  & 0.068  & 0.102 \\
Only minor powers			& -0.647 & 0.178 & -0.618  & 0.078  & 0.188  & 0.344 \\
Constant 					& -0.128 & 0.536 & -0.181  & 0.211  & 0.562  & 0.840 \\
\hline
\multicolumn{7}{|l|}{$N$ = 39,988 dyad-year observations for 1,178 dyads and 146 units.} \\
\hline
\end{tabular}
\end{footnotesize}
\label{tab:russett} 
\end{table}

We attempted also to study how inferences might change with a random effects approach and a spatial error lag approach.  The appropriate random effects model would be a three-way random effects model to account for each state in a dyad as well as the dyad as a whole over time.  We attempted to fit three-way random effects logit models using the {\tt lmer} package in {\tt R} and the {\tt melogit} function in Stata.  We posted the jobs to a university cluster with 24 Intel Xeon cores and 48 gigabytes of RAM per node.  In neither case did the random effects models converge. In fact, in neither case could we obtain estimates of any form after days on the cluster.  For the spatial error lag model, we sought to implement both the spatial error lag model in the {\tt spdep} package in {\tt R} and the two-step estimator of  \citet{neumayer-plumper-2010-spatial} in Stata.  The latter take as an input dyadic dependence information that can be produced using the {\tt spundir} package in Stata.  The spatial error lag model in the {\tt spdep} failed to produce estimates and {\tt spundir} package failed to resolve within reasonable time (a week) while running on the same cluster.\footnote{See also footnote \ref{fn:sel-model-probs}.  The computational complexity is due to the fact that the model attempts to account for dependence across all countries in the dataset (rather than small subsets, as would be the case in a typical spatial analysis) and the fact that in this application the set of cross-section units changes over time and therefore does not allow for a stable adjacency matrix over time.}  This brings to the foreground the issue of computational complexity, by which our cluster robust estimator is much less demanding.

\section{Conclusion}

We have established convergence properties for a non-parametric variance estimator for regression coefficients that accounts for dyadic clustering. The estimator applies no restrictions on the dependency structure beyond the dyadic clustering assumption.  The estimator is robust to the regression model being misspecified for the conditional mean. Such robustness is important because regression analysis typically relies on linear (in coefficients) approximations of unknown conditional mean functions \citep{angrist_pischke09, buja-etal2014-models-as-approximations, hubbard-etal2010-to-gee}. Of course our analysis in no way excuses analysts from their responsibility to obtain as good an approximation as possible nor does a robust fix for standard errors also solve the problem of finding a good approximation. Given a reasonable approximation for the conditional mean, the methods we have developed here allow for more accurate asymptotic statistical significance tests and confidence intervals.

The estimator is consistent in the number of units that form the basis of dyadic pairs. Simulations show that the estimator approaches the true standard error with modestly-sized samples from a reasonable data generating process. Applications show that inferences can be seriously anti-conservative if one fails to account for the dyadic clustering. This estimator is a natural complement to the non-parametric and semi-parametric regression analyses that are increasingly common in the social sciences \citep{angrist_pischke09}.  Given that we can express the estimator as the sum of simpler and easy-to-compute robust variance estimators, it could be applied to any estimator for which a cluster-robust option is available.

A cost of the robust approach is efficiency.  Our simulations show that a two-way (for cross-sectional data) random effects model can be considerably more efficient and provide reliable inference when the conditional mean is correctly specified.  Our robust estimator could provide the basis of a test for misspecification \citep{king-roberts2014-robust, white80, white-1981-detection-misspec}.  If there is little evidence of misspecification, the random effects estimator would be a reasonable choice given its efficiency.  However, problems of computational non-convergence may make the random effects estimator infeasible---we encountered this in our application when we tried to use a three-way random effects model for dyadic panel data.

Of course, accounting for dynamic dyadic clustering may fail to fully account for all relevant dependencies in the data.  For example, units may exhibit higher-order network effects: for example a shock to unit $A$ may affect unit $C$ through connections that run via a third unit, $B$. In such cases, the methods developed here will likely yield standard errors that are too small, although they should still outperform methods that fail even to account for dyadic clustering.
 
\clearpage
\appendix

\section{Proof of Lemma \ref{lemma-var}}

\begin{proof}
We have 
$$
\sqrt{N}(\hat \beta - \beta) = \left[\frac{1}{D} \sum_{d=1}^D X_d X_d' \right]^{-1} \frac{\sqrt{N}}{D} \sum_{d=1}^D X_d \epsilon_d.
$$
Take $[(1/D) \sum_{d=1}^D X_d X_d' ]^{-1}$ first. By continuity of the inverse, the finite support of $X_d$ and $e_d$, $\X$ full rank, and the fact that the sum of dependent terms in this sum is $O(D^{3/2})$ \citep[Eq. 2.8.4]{lehmann1999}, we have that 
$$
\left[\frac{1}{D} \sum_{d=1}^D X_d X_d' \right]^{-1} \overset{p}{\rightarrow} \E[X_d X_d']^{-1}.
$$
By Slutsky's theorem, $\sqrt{N}(\hat \beta - \beta)$ has the same asymptotic distribution as $\E[X_d X_d']^{-1}\frac{\sqrt{N}}{D}\sum_{d=1}^D X_d\epsilon_d$, which has mean zero and variance,
\begin{equation}
V = \frac{N}{D^2}\E[X_dX_d']^{-1}\Var\left[\sum_{d=1}^D X_d\epsilon_d\right]\E[X_dX_d']^{-1}. \label{varB}
\end{equation}

Then,
\begin{align}
\Var\left[\sum_{d=1}^D X_d\epsilon_d \right] & = \sum_{d=1}^D \sum_{d'=1}^D \Cov[X_d\epsilon_d, X_{d'}\epsilon_{d'}]\nonumber \\
& = \sum_{d=1}^D\sum_{d'=1}^D \E\{X_{d}X'_{d'} \Cov[\epsilon_d,\epsilon_{d'}|\X]\}\nonumber \\
& = \sum_{d=1}^D \left( \underbrace{\E\{X_{d}X'_d \Var[\epsilon_d|\X]\}}_{A} + \underbrace{\sum_{d' \in \mathcal{S}(d)} \E\{X_{d}X'_{d'} \Cov[\epsilon_d,\epsilon_{d'}|\X]\}}_{B} \right) \label{varxe}
\end{align}
where $\mathcal{S}(d) = \{d' \ne d \} \cap \{d': s(d)\cap s(d') \ne \emptyset \}$, the set of dyads other than $d$ that share a member from $d$.
\end{proof}

\section{Proof of Proposition \ref{prop1}}

\begin{proof}
%Convergence in expectation follows from consistency of $\hat \beta$. To establish convergence in probability, we need only establish that $\Var[N \hat V] \arrowp 0$ as $N \rightarrow \infty$. 
By Chebychev's inequality, two conditions are sufficient for $N\hat V - V \overset{p}{\rightarrow} 0$ as $N \rightarrow \infty$: (i) $\E[\hat V - V] \rightarrow 0$, and (ii) $\Var[N\hat V - V] = \Var[N\hat V] \rightarrow 0$ \citep[Thm. 2.1.1]{lehmann1999}. 

For (i), consider the interior (``meat'') of $\hat V$:
$$
\hat \Sigma = \sum_{d=1}^D \left( X_{d}X'_d e_d^2 + \sum_{d' \in \mathcal{S}(d)} X_{d}X'_{d'} e_d e_{d'}\right)
$$ 
and the corresponding term for $V$:
$$
\Sigma = \sum_{d=1}^D \left( \E\{X_{d}X'_d \Var[\epsilon_d|\X]\} + \sum_{d' \in \mathcal{S}(d)} \E\{X_{d}X'_{d'} \Cov[\epsilon_d,\epsilon_{d'}|\X]\} \right).
$$ 
We have
$$
\E[\hat \Sigma] = \sum_{d=1}^D \left( \E\{X_{d}X'_d \E[ e_d^2|\X]\} + \sum_{d' \in \mathcal{S}(d)} \E \{X_{d}X'_{d'} \E[e_d e_{d'}|\X]\}\right) \rightarrow \Sigma
$$
by consistency of $\hat \beta$, an implication of Lemma \ref{lemma-var}. Lemma \ref{lemma-var} also established that $(\X'\X/D)^{-1} \overset{p}{\rightarrow} \E[X_d X_d']^{-1}$, while boundedness of $X_d$ implies uniform boundedness of $\E[X_d X_d']^{-1}$.  Taking these elements together, uniform continuity implies $\E[\hat V - V] \rightarrow 0$ (cf. \citealp{white80}, Thm. 1). 

For (ii), we follow the strategy of \citep[section 2.8]{lehmann1999}, which requires showing that growth of covariance contributions due to dependent units is bounded in $N$. Begin by rewriting $N \hat V$ as
\begin{align*}
N \hat V = N(\X'\X/D)^{-1}\left(D^{-2} \sum_{d} \sum_{d'} I_{d,d'}X_{d}X'_{d'} e_d e_{d'}\right)(\X'\X/D)^{-1}
\end{align*}
where $I_{d,d'}$ is an indicator function denoting a pair of dyads $d$ and $d'$ that share a member, as defined above. We have established that $(\X'\X/D)^{-1}$ converges in probability to a finite limit, $E[X_dX_d']^{-1}$, so we can treat this as $O(1)$. Given bounded support, $\Var[N \hat V]$ is of the same order as
\begin{align} \Omega &= N^2 D^{-4} \Var\left[ \sum_{d} \sum_{d'} I_{d,d'}X_{d}X'_{d'} e_d e_{d'}\right] \nonumber \\
& = N^2 D^{-4} \sum_{d} \sum_{d'} \sum_{d''} \sum_{d'''} \Cov(I_{d,d'}X_{d}X'_{d'} e_de_{d'}, I_{d'',d'''}X_{d''}X'_{d'''} e_{d''}e_{d'''}). \nonumber
\end{align} 
To show $\Var[N\hat V] \rightarrow 0$, it is sufficient to establish that $\Omega$ is at most $O(N^{-1})$. Consider quadruples of dyads indexed by $d, d', d'', d'''$ such that $d \neq d' \neq d'' \neq d'''$. Suppose there are $Q$ quadruples and define $p_Q(A) = |A|/Q$.  The covariance terms take positive values when $ I_{d,d'} = I_{d'',d'''}=1$, which occurs at rate $O(N^{-2})$ (that is, when we have pairs of terms that are actually summed into $\hat V$), and we have dependence across $(d,d')$ and $(d'',d''')$. Then, the six distinct ways that such dependence can occur and the associated proportion of quadruples for which these occur for a sample of size $N$ are as follows:
\begin{align*}
& p_Q(\{d \ne d'' \} \cap \{d'': s(d)\cap s(d'') \ne \emptyset \}) = O(N^{-1}), \\
& p_Q(\{d \ne d''' \} \cap \{d''': s(d)\cap s(d''') \ne \emptyset \}) = O(N^{-1}), \\
& p_Q(\{d \ne d'' \ne d'''\} \cap \{d'', d''': s(d)\cap s(d'') \ne \emptyset \cap s(d)\cap s(d''') \ne \emptyset\})= O(N^{-2}), \\
& p_Q(\{d' \ne d'' \} \cap \{d'': s(d')\cap s(d'') \ne \emptyset \})=O(N^{-1}), \\
& p_Q( \{d' \ne d''' \} \cap \{d''': s(d')\cap s(d''') \ne \emptyset \})=O(N^{-1}), \text{ and } \\
& p_Q(\{d' \ne d'' \ne d''' \} \cap \{d'', d''': s(d')\cap s(d'') \ne \emptyset \cap s(d')\cap s(d''')\ne \emptyset \}) = O(N^{-2}). 
\end{align*}
Therefore the proportion of quadruples yielding a positive covariance is the proportion of cases contributing non-zero values to the sum
$\Omega$. This proportion is bounded by the proportion of cases where the indicator
functions are both one and the order of the proportion of dependent
quadruples from across the six cases characterized above, that is $O(N^{-2}N^{-1}) = O(N^{-3})$. Then $\Var[N\hat V]$ is at most $O(N^2 D^{-4} D^4 N^{-3}) = O(N^{-1}) \rightarrow 0$.

%The proportion of cases contributing non-zero values to the sum,
%$\Omega$, is bounded by the proportion of cases where the indicator
%functions are both one and the order of the proportion of dependent
%quadruples from across the six cases characterized above, that is
%O(N-2N-1) = O(N-3).

\end{proof} 

\section{Proof of Proposition \ref{prop-decomp}}

\begin{proof}
Consider the interior (``meat'') of $\hat V_r$:
$$
\hat \Sigma_r = \sum_{d=1}^D\left(X_d'e_de_d'X_d + \sum_{d' \in \mathcal{S}(d)} X_d'e_de_{d'}'X_{d'} \right) = \sum_{d=1}^D X_d'e_de_d'X_d + \sum_{d=1}^D \sum_{d' \in \mathcal{S}(d)} X_d'e_de_{d'}'X_{d'}.
$$ 
Taking the second component,
$$
\sum_{d=1}^D \sum_{d' \in \mathcal{S}(d)} X_d'e_de_{d'}'X_{d'} = \sum_{i=1}^N C_i, \text{ with }  C_i = \sum_{j\ne i} \sum_{k \ne i,j} X'_{d(i,j)}e_{d(i,j)}e'_{d(i,k)}X_{d(i,k)}.
$$
Define 
$$
D_i = \sum_{j \ne i} X'_{d(i,j)}e_{d(i,j)}e'_{d(i,j)}X_{d(i,j)}
 \text{\hspace{1em} and \hspace{1em}} E_i = \sum_{j,k \ne i} \sum_{t=1}^{T(d(j,k))} X_{d(j,k)t}X'_{d(j,k)t}e^2_{d(j,k)t}.
$$
Then working with $\hat \Sigma_{C,i}$, $\hat \Sigma_D$, and $\hat \Sigma_0$ as defined in the proposition, 
$$\hat \Sigma_{C,i} = C_i + D_i + E_i \text{, \hspace{1em}} 2 \hat \Sigma_D = \sum_{i=1}^N D_i\text{, \hspace{1em} and \hspace{1em}} (N-2) \hat \Sigma_0 = \sum_{i=1}^N E_i.
$$
Then
\begin{align}
\hat \Sigma_r & = \hat \Sigma_D + \sum_{i=1}^N C_i \nonumber \\
& = \hat \Sigma_D + \sum_{i=1}^N [C_i + D_i + E_i  - D_i - E_i] \nonumber \\ 
& = \sum_{i=1}^N \hat \Sigma_{C,i} - \hat \Sigma_D - (N-2) \hat \Sigma_0, \nonumber 
\end{align}
and by linear distributivity for $(\X'\X)^{-1}$ 
$$
\hat V_r = \sum_{i=1}^N \hat V_{C,i} - \hat V_D - (N-2)\hat V_0,
$$
with $\hat V_{C,i}$, $\hat V_D$, and $\hat V_0$ as defined in the proposition.
\end{proof}

\section{Weighted Observations}

Weighting is a common way to adjust for unequal probability sampling of dyadic interactions, among other applications.  The extension to the weighted case is straightforward.  Assume weighted directed dyad observations with weights finite and fixed; denote the weight for dyad $d$ as $w_d$. Define the sample data as $\X = (X'_1...X'_D)'$ and $Y = (Y_1 ...Y_D)'$ and the matrix of weights as $\W=\text{diag}\{w_1, \hdots, w_D\}$. Then the weighted least squares estimator is $$ \hat \beta_w =(\X' \textbf{W} \X)^{-1} \X' \textbf{W}Y. $$

\begin{corollary}
For the weighted dyads case, assume the same conditions as in Proposition \ref{prop1} and consider the following variance estimator

\begin{equation}
\hat V_w = (\X' \W \X)^{-1} \sum_{d=1}^D \left(X_{d}X'_d w_{d}^{2}e_d^{2} + \sum_{d' \in \mathcal{S}(d)}X_{d}X'_{d'} w_{d}w_{d'}e_d e_{d'} \right)(\X' \W\X)^{-1}.
\end{equation}

Then as $N \rightarrow \infty$, 
$$
N\Var[\hat \beta_w - \beta_w] - V_w \overset{p}{\rightarrow} 0,$$
and
$$
N\hat V_w - V_w \overset{p}{\rightarrow} 0,
$$
where 
$$
V_w = \frac{N}{D^{2}} \E[w_{d}X_d X_d']^{-1}
\E\left( \E\left[X_dX'_d w_{d}^{2}\Var[\epsilon_d|\X]\right] + \sum_{d' \in \mathcal{S}(d)}\E\left[X_dX'_{d'}w_{d}w_{d'}\Cov[\epsilon_d, \epsilon_{d'}|\X]\right]\right)
\E[w_{d}X_dX_d']^{-1}.
$$
\end{corollary}

\section{Generalized Linear Models}

An implementation for generalized linear models follows the usual $M$-estimation results (\citealp{stefanski_boos2002_m_estimation}; \citealp{wooldridge10_book}, Ch. 12). Given an estimating equation, $\psi(D;\theta)$, on data $D$, and with parameters $\theta$ and parameter estimates $\hat \theta$, the sandwich approximation for the variance is $\A^{-1} \B \{\A^{-1}\}'$, where
$$
\A = \E\left[-\frac{\partial}{\partial \theta'}\psi(D;\hat \theta)\right] \text{\hspace{1em} and \hspace{1em}}\B = \E \psi(D;\hat \theta)\psi(D_i;\hat \theta)'.
$$
For logistic regression the estimating equation is,
$$
\psi((\X_d, Y_d); \hat\beta) =  \X_d(Y_d - p_d),
$$
where $p_d = \text{expit}(\X_d\hat\beta)$, the predicted probability for dyad $d$. The plug-in variance estimator for logistic regression coefficients is given by
$$
\hat V_{l} = (\X'\M\X)^{-1}\sum_{d=1}^D\left(X_dX_d'r^2_d + \sum_{d' \in \mathcal{S}(d)} X_dX_d'r_d r_{d'} \right)(\X'\M\X)^{-1}
$$
where 
$$
\M = \text{diag}(p_d(1-p_d))_{D \times D} 
\hspace{.5em} \text{ and } \hspace{.5em}  r_d = Y_d - p_d.
$$
The extensions to repeated and weighted observations follow analogously.

\clearpage
\singlespace
%\bibliography{/Users/cyrussamii/Dropbox/bib/all}
%\bibliography{all}
%\bibliography{/Users/nlsamii/Dropbox/bib/all}

\clearpage
\setcounter{page}{1}

\begin{center}
{\Large Cluster Robust Variance Estimation for Dyadic Data}\\
Supporting Information Not for Publication
\end{center}

\clearpage
\section*{Additional Simulation Evidence}

\begin{figure}[!hp]
\centering
\includegraphics[width=1\textwidth]{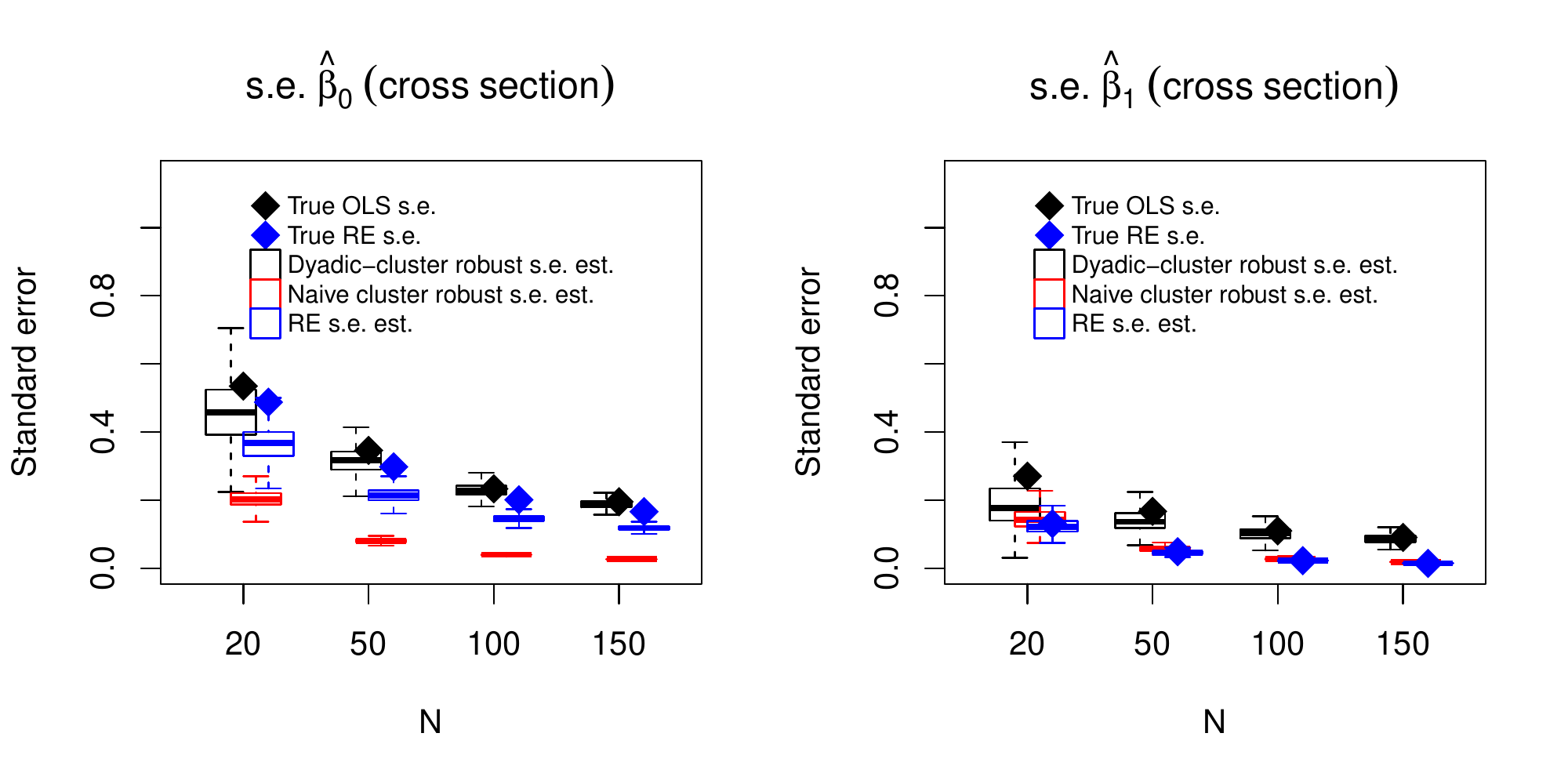}
\caption{Normal shocks case (same as main text).}
\label{fig:norm}
\end{figure}

\begin{figure}[!hp]
\centering
\includegraphics[width=1\textwidth]{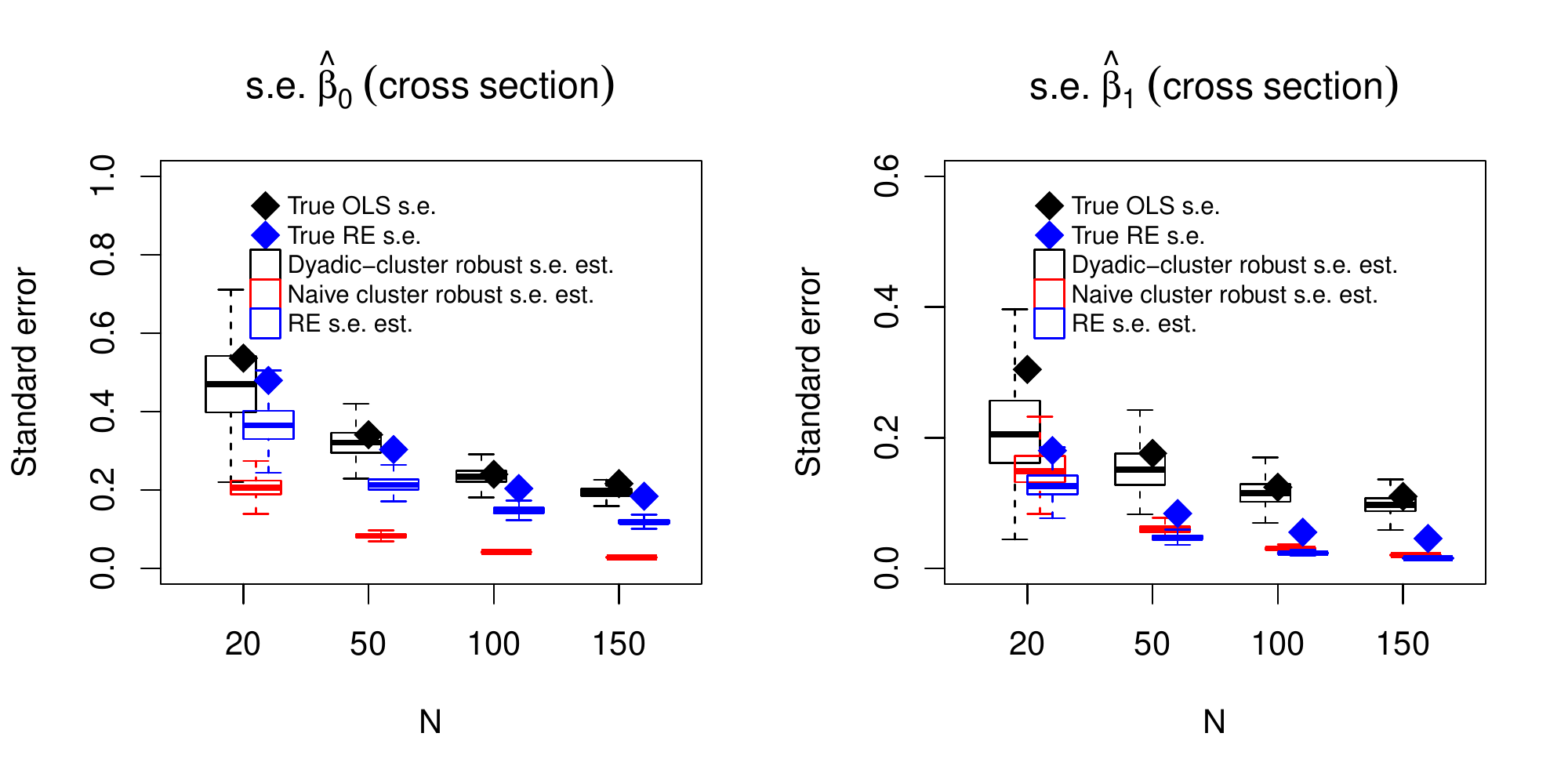}
\caption{Misspecification case.}
\label{fig:miss}
\end{figure}

Figure \ref{fig:norm} and \ref{fig:miss} show additional simulation results where we make comparisons to mixed effects models---specifically, to a two-way random effects model that incorporates random effects for each of $N$ units in the data.  All random effects models were fit using the {\tt lmer} package in {\tt R}.  In both sets of graphs, black box plots show the distribution of standard error estimates for OLS coefficients using the proposed dyadic cluster robust estimator. Gray box plots show standard error estimates for OLS coefficients from a heteroskedasticity robust estimator.  Blue box plots show the distribution of standard error estimators from a two-way random effects model with a random effect for each of $N$ units. Black diamonds show the true standard error of the OLS coefficients and the blue diamonds show the true standard error of the random effects coefficients.

Figure \ref{fig:norm} uses the same data generating process as was described in the main text, and presents the same results for OLS with either the dyadic cluster robust standard error estimator or the heteroskedastic robust estimator.  In addition, we have added random effects estimates.  Given that the data generating process conforms exactly to the assumptions of a two-way random effects model, we obtain the straightforward result that the random effects estimates are both more efficient (the blue diamonds tend to be lower than the black ones) and the random effects standard errors are consistent. 

\begin{figure}[!hp]
\centering
\includegraphics[width=.5\textwidth]{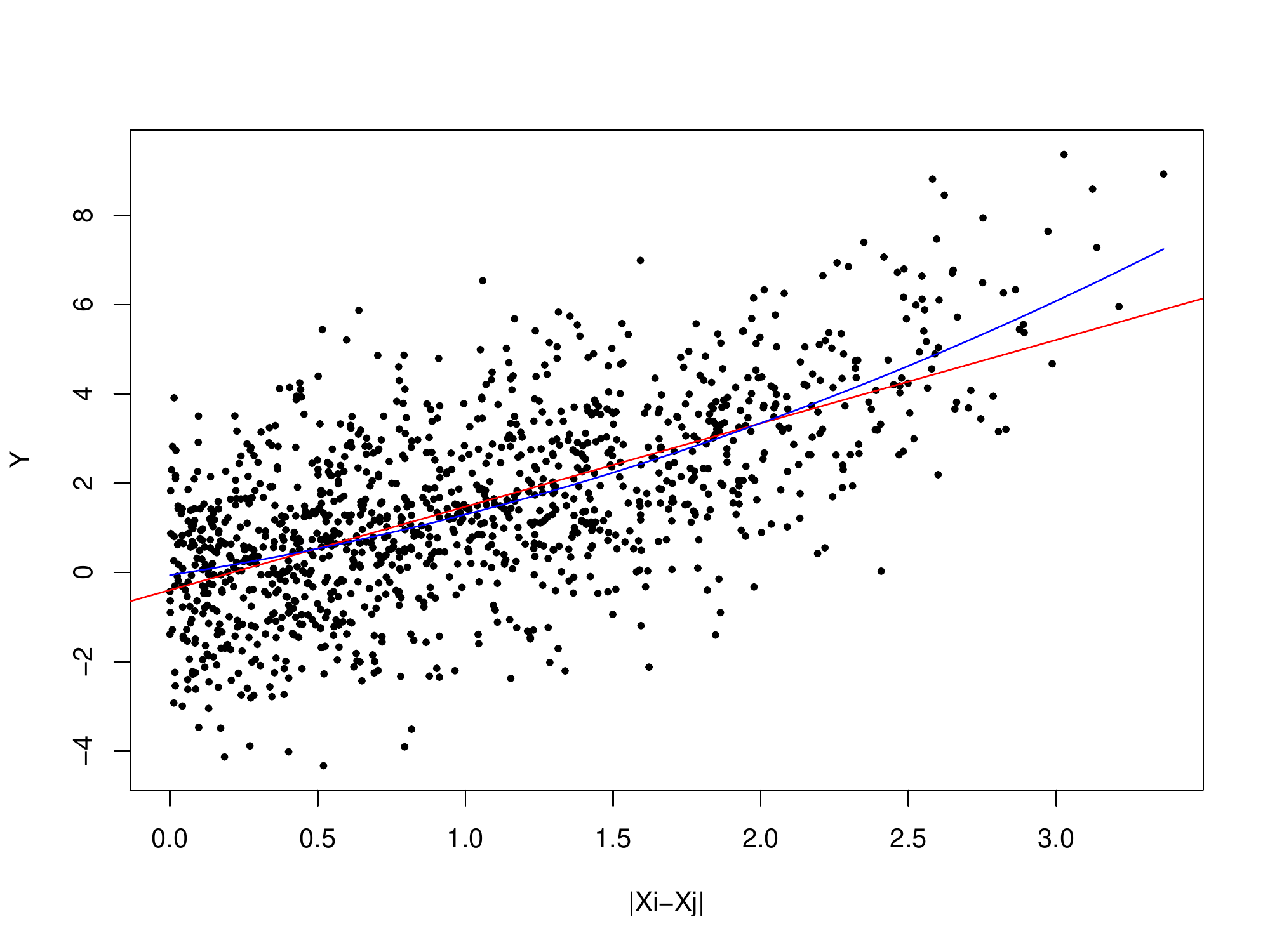}
\caption{Illustrative scatter plot and linear approximation for the misspecification case.  The blue line is the true conditional mean, and the red line is the linear approximation.}
\label{fig:miss-scatter}
\end{figure}

Figure \ref{fig:miss} illustrates the robustness of the dyadic cluster robust estimator relative to a random effects estimator. In this case, we induce mild misspecification by assuming that the true data generating process is
$$
Y_{d(i,j)t} = \beta_0 + \beta_1 |X_i - X_j| + \beta_2 (X_i - X_j)^2 + \underbrace{\alpha_i + \alpha_j + \nu_{d(i,j)t}}_{\epsilon_{d(i,j)t}},
$$ 
where all parameters are set as in the main text but we also have $\beta_2 = 0.25$.   We nonetheless fit a model that includes only $|X_i - X_j|$, thereby mildly misspecifying the model such that we do not account for the nonlinearity induced by the $\beta_2(X_i - X_j)^2$ term.  Figure \ref{fig:miss-scatter} shows the scatter plot of $Y_{d(i,j)t}$ against $|X_i - X_j|$, the linear approximation (in red) for one of the simulation replicates ($N=50$), and the true conditional mean (in blue).    Such linear approximations are the norm in social science research \citep{buja-etal2014-models-as-approximations}.  In this case, we see that, despite the mild misspecification, the dyadic cluster robust standard error correctly characterizes the standard error of the linear approximation coefficients (the black box plots still zero in on the black diamonds).  The random effects model is still more efficient (the blue diamonds still tend to be below the black diamonds).  But the random effects standard error estimator (as computed in the {\tt lmer} package for {\tt R}) is inconsistent and anti-conservative: the blue box plots converge to a limit that is substantially below the true standard error (the blue diamonds).  

\clearpage
\section*{Speed Dating Application}
Our second application is based on \cite{Fisman2006}'s seminal study of the determinants of mate selection in a ``speed-dating'' experiment. We replicate their primary analysis for female participants (as reported in column 1 of their Table III) using data from 21 dating sessions conducted by the experimenters.\footnote{The dataset to which we had access is not identical that used in the original \cite{Fisman2006} paper. Rather, it is the version made available in association with \citet[pp. 322-323]{gelman-hill2007-multilevel-book}, available at \url{http://www.stat.columbia.edu/~gelman/arm/software}. We contacted the original study authors and they indicated that the dataset from the original paper was not available.\label{fn:fisman-data}} These data include 278 women paired into 3,457 female-male dyads. \cite{Fisman2006} regress a binary indicator for whether the female subject desired contact information for a male partner on the subject's ratings of the partner's ambition, attractiveness, and intelligence based on a 10-point Likert scale.  The regression controls for female-subject fixed effects and weights observations contributed by each female subject by the inverse of the number of partners with which she was paired.  Table \ref{tab:fisman} shows the results of our re-analysis.\footnote{For the reasons explained in footnote \ref{fn:fisman-data}, our estimates differ slightly from those that appear in the original paper.}  Again, we see that accounting for the dyadic clustering (column 4) yields standard error estimates that are larger than what one would get if one ignored all clustering (column 2), although in this case there are no pronounced differences with what one gets when one only accounts for within-subject clustering (column 3). This is to be expected given that the amount of dyadic dependence in this dataset is limited: dyads were formed only {\it within} sessions that included between 5 and 22 male partners for each female subject.  In addition, the female subjects are never paired with each other.  The only clustering that occurs is therefore within female subjects and then for male partners that appear across multiple female subjects.  The former is addressed in part by the female subject-specific fixed effects. The latter is limited by the within-session pairings.

\begin{table}[!h]
\caption{Estimates and robust standard error estimates from a linear fixed-effects regression of an indicator for mate selection between dyads on dyad-level predictors, based on \citet[Table III, Column 1]{Fisman2006}.} 
\centering
\begin{footnotesize}
\begin{tabular}{|r|c|ccc|}
\hline
& (1) & (2) & (3) & (4)\\
&  & Het.  & Naive Cluster& Dyadic Cluster \\
Predictor of Mate Selection & Coef. & Robust S.E. & Robust S.E. & Robust S.E. \\
\hline
Ambition &          	0.0192 & 0.0052 & 0.0057 & 0.0061 \\
Attractiveness   &       0.1157  & 0.0041 & 0.0051 & 0.0054 \\
Intelligence &        	0.0465  & 0.0062 & 0.0073 & 0.0074 \\
\hline
\multicolumn{5}{|l|}{$N$ = 3,457 dyad observations for 278 subjects.} \\
\hline
\end{tabular}
\end{footnotesize}
\label{tab:fisman} 
\end{table}

\end{document}